\documentclass[12pt]{article}
\usepackage{fancyhdr}
\usepackage{tocbibind}
\usepackage[titletoc,title]{appendix}
\usepackage{amsmath}
\usepackage{amssymb}
\usepackage[usenames,dvipsnames]{color}
\usepackage{bbold}
\usepackage{tensor}
\usepackage{parskip}
\usepackage{graphicx}
\usepackage{hyperref}
\usepackage[affil-it]{authblk}
\advance \topmargin by -\headheight
\advance \topmargin by -\headsep   
\evensidemargin \oddsidemargin
\newcommand*{\dt}[1]{\widetilde{\widetilde{#1}}}
\newcommand*\dd{\mathop{}\!\mathrm{d}}
\newcommand{\Lagr}{\mathcal{L}}
\newcommand{\beq}{\begin{equation}}
\newcommand{\eeq}{\end{equation}}
\newcommand{\mcal}{\mathcal}
\newcommand{\ket}[1]{\left|  #1 \right\rangle}
\newcommand{\braket}[1]{\ensuremath{\left\langle #1 \right\rangle}}

\definecolor{deepblue}{rgb}{0.2,0.2,0.5}

\def\simgt{\mathrel{\lower2.5pt\vbox{\lineskip=0pt\baselineskip=0pt
           \hbox{$>$}\hbox{$\sim$}}}}
\def\simlt{\mathrel{\lower2.5pt\vbox{\lineskip=0pt\baselineskip=0pt
           \hbox{$<$}\hbox{$\sim$}}}}

\fancypagestyle{firststyle}
{
   \fancyhead[RO]{SLAC-PUB-16383}
   \fancyfoot[C]{\footnotesize  \thepage\ }
}

\begin{document}
\title{\bf\Large{Extended Conformal Symmetry in $d\neq 4$ :\\
 Conformal Symmetry of Abelian Gauge Theory \\in the Physical Sector} }
\author[1]{Kelly Yu-Ju Chiu\thanks{yujuchiu@stanford.edu}}
\author[1]{Stanley J. Brodsky\thanks{sjbth@slac.stanford.edu}} 
\affil[1]{\normalsize{SLAC National Accelerator Laboratory,

 Stanford University, Stanford, California 94039, USA}} 
\renewcommand\Authands{ and } 
%\author{{Kelly Yu-Ju Chiu} }
%\author{{Stanley J. Brodsky}}
\setlength{\affilsep}{1em}

%\affil{\normalsize{SLAC National Accelerator Laboratory,\\
% Stanford University, Stanford, California 94039, USA}}
\date{}

\maketitle
\abstract{Abelian gauge theory in $d\neq 4$ spacetime dimensions is an example of a scale invariant theory which does not possess conformal symmetry -- the special conformal transformation(SCT) explicitly breaks the gauge invariance of the theory. In this work, we construct a non-local gauge-invariant extension of the SCT, which is compatible with the BRST formalism and defines a new symmetry  of the physical Hilbert space of the Maxwell theory for any dimension $d\geqslant 3$. We prove the invariance of the Maxwell theory in $d\geqslant 3$ by explicitly showing that the gauge-invariant two-point correlation functions, the  action, and the classical equation of motion are unchanged under such a transformation.}
\thispagestyle{firststyle} 
\cleardoublepage 
\section{Introduction}
Conformal field theories (CFT) are of great interest in physics research since conformal symmetry appears to be an underlying fundamental property of nature. In a second-order phase transition, the divergence of correlation length  implies that a theory is scale invariant at the critical point. In fact,
the theory at the critical point almost always possesses a larger symmetry than scale invariance -- the conformal symmetry. In perturbative string theory, conformal invariance of the worldsheet action is crucial for consistency of the theory. The AdS/CFT correspondence provides new insights into the non-perturbative domain of quantum field theories, such as QCD, thus also promotes the significance of CFT \cite{Brodsky:2014yha}. 

CFTs have been studied extensively in $d=2$ spacetime dimensions due to the powerful conformal group algebra specific to $d=2$. However, the operation of conformal transformations in $d \geqslant 3$ has not been fully understood. For instance, the relation of scale invariance and conformal invariance is unclear in general dimensions. In $d=2$, Zamolodchikov and Polchinski showed that scale invariance guarantees conformal invariance in a unitary theory \cite{Zamolodchikov:1986gt,Polchinski:1987dy}. For $d \geqslant 3$, unitary theories which exhibit scale invariance usually are conformally invariant; however, a general proof that scale implies conformal invariance has not been discovered. In fact, a few counterexamples are known \cite{Fortin:2011ks,Iorio:1996ad,ElShowk:2011gz}. 

One simple counterexample of a unitary scale invariant theory which is not conformally invariant in general dimensions is the free Maxwell theory, where
\beq
\Lagr(x)= -\frac{1}{4}F^{\mu\nu}(x)F_{\mu\nu}(x).
\eeq
This theory is known to break conformal invariance in $d \neq 4$ due to the non-vanishing trace of the stress-energy tensor. Interestingly, El-Showk \textit{et.al} argued that the conformal invariance of the Maxwell theory in $d\neq 4$ can be restored in a specific $\xi$-gauge with $\xi=d/(d-4)$ using the Faddeev-Popov procedure \cite{ElShowk:2011gz}. Nevertheless there is a difficulty in this construction: the special conformal transformation (SCT) maps physical states to zero-norm states which are unphysical in the BRST Hilbert space. Thus, the conformal invariance in this theory cannot be understood as a property of the physical sector of the theory alone, \textit{i.e.} the unitary gauge invariant Maxwell theory. 

On the other hand, El-Showk \textit{et.al}  showed that the two-point correlation function of the field strength $F^{\mu\nu}$ is conformally invariant. Evidently the correlation function of the field strength is gauge invariant, so if it is SCT invariant under a certain $\xi$-gauge, it should also be invariant in any gauge. It is thus intriguing to ask, if the original conformal invariance is not present in the physical Maxwell theory, why is the physical correlation function still invariant under such transformations? 

In this paper, we define a non-local gauge-invariant extended special conformal transformation (ESCT), in which the field strength transforms in a similar but gauge invariant way. It is compatible with the BRST formalism, and only maps between states within the physical Hilbert space, in contrast to the usual SCT \cite{Deser:1994ca}. In fact, we shall demonstrate that the operation of the ESCT reveals a symmetry of the free Maxwell theory in $d\geqslant 3$ by showing that the action, the classical equation of motion(EOM), and physical two-point correlation functions are all invariant under the ESCT. 

The paper is organized as follows: In Section 2, we briefly remind the readers how conformal transformations act both on coordinates and fields in $d\geqslant 3$. In Section 3, we review why the  classical Maxwell theory is not conformally invariant. We then move on to the quantum version, and by studying the BRST Hilbert space, we show that the SCT is not compatible with gauge invariance of the theory since it mixes physical states with the unphysical ghost degrees of freedom. In Section 4, we employ a method of decomposing the gauge field and explicitly  compute how the SCT operator acts on physical and unphysical components. 

It is important to avoid mixing between physical and unphysical degrees of freedom present induced by the SCT, thus in Section 5, following the gauge field decomposition, we define an ESCT which commutes with the BRST generator and only acts within the physical sector. We then prove the ESCT invariance of both the classical and quantum Maxwell theory. Finally, we conclude in Section 6.

\section{Conformal Transformation}

An infinitesimal coordinate transformation $x^\mu \rightarrow x'^\mu=x^\mu + \epsilon^\mu(x)$ generated by the connected part of the conformal group preserves the metric up to a scale factor $\lambda (x)$:  
$g'_{\mu\nu}(x') = \lambda(x) g_{\mu\nu} (x)$, where $g_{\mu\nu}=\text{diag}(1,-1,-1,...)$ is the flat spacetime metric.  

By requiring that a coordinate transformation be conformal, we must have  \cite{DiFrancesco1997}   
\beq
\partial_\mu \epsilon_\nu + \partial_\nu \epsilon_\mu = \frac{2}{d} g_{\mu\nu} \partial \cdot \epsilon.
\eeq

It can be shown that for  $d\geqslant 3$,  only terms up to $x^2$ are allowed in $\epsilon(x)$. There are four classes of solutions\cite{Rychkov}:
\\
\begin{itemize}
\item  Translation :  
$\epsilon^\mu(x)= a^\mu$, where $a^\mu$ is a constant. 
%x^\mu+  a_\alpha \,\delta^\alpha x^\mu$, $ \delta^\alpha x^\mu= g\indices{^\alpha ^\mu}.
\\
\item  Lorentz transformation : 
$\displaystyle \epsilon^\mu (x)= \omega\indices{^\mu_\nu} \, x^\nu$, $\omega\indices{^\mu^\nu}= -\omega\indices{^\nu^\mu}$.  
%= x^\mu+ \frac{1}{2} \omega_{\alpha \beta}\, \delta\indices{^\alpha^\beta}x^\mu$, $ \delta\indices{^\alpha^\beta}x^\mu = (g^{\mu\alpha}g^{\nu\beta} - g^{\mu\beta}g^{\nu\alpha} ) x_\nu $,  $\omega\indices{^\mu^\alpha}= -\omega\indices{^\alpha^\mu}$.
\\
\item  Scaling (dilatation) : 
$\displaystyle \epsilon^\mu(x)= \sigma x^\mu$.
%x^\mu + \sigma \, \delta x^\mu$, $\delta x^\mu = x^\mu 
\\
\item  Special conformal transformation (SCT) : 
the SCT can be understood as the series of operations (inversion $\rightarrow$ translation $\rightarrow$ inversion ) : 
\beq
x^\mu \rightarrow \frac{x^\mu}{x^2} \rightarrow \frac{x^\mu}{x^2} -c^\mu \rightarrow \frac{\frac{x^\mu}{x^2} -c^\mu} {(\frac{x^\mu}{x^2} -c^\mu)^2} =x'^\mu.
\eeq  
Expanding the result to first order, we obtain the infinitesimal SCT  
\beq
\epsilon^\mu(x)= 2 c \cdot x \, x^\mu - c^\mu x^2,
\eeq
and 
\beq
\displaystyle \frac{\partial x'^\mu}{\partial x^\nu}= (1 +2 c \cdot x )\: \delta \indices{^\mu_\nu} + (2x^\mu c_\nu - 2x_\nu c^\mu).
\eeq
Thus, the SCT can be interpreted as a combination of a position-dependent scaling and a Lorentz rotation.  

For convenience, for the SCT, we define 
\beq
\delta^\sigma x^\mu \equiv \frac{\delta \epsilon^\mu(x)}{\delta c_\sigma}=2x^\sigma x^\mu - g^{\sigma \mu} x^2.
\eeq
\end{itemize} 

In $d\geqslant 3$, translation, rotation, dilatation, and the SCT together form the $SO(d,2)$ conformal group.
In addition, one can express an infinitesimal conformal transformation as: 
\beq
 \frac{\partial x'^\mu}{\partial x^\nu}= \delta \indices {^\mu_\nu } + \sigma(x) \delta \indices {^\mu_\nu } + \omega \indices {^\mu_\nu} (x) ,
 \eeq
where 

\beq
\sigma(x)=  \begin{cases} 0  & \text{translation and rotation}  \\\\ 
 \sigma  & \text{dilatation }  \\\\ 
 2 c \cdot x  & \text{SCT }  \end{cases}
 \label{trans1}
\eeq
and  
\beq
\omega \indices {^\mu_\nu } (x)=  \begin{cases} 0  & \text{translation and dilatation}  \\\\ 
\omega  \indices {^\mu_\nu }  & \text{rotation }  \\\\ 
2x^\mu c_\nu - 2x_\nu c^\mu  & \text{SCT }  \end{cases}
\label{trans2}
\eeq
 
With the appropriate $\sigma(x)$ and $\omega_{\mu\nu}(x)$ as defined in   (\ref{trans1}) and (\ref{trans2}) for each transformation, a conformal transformation acts on a field with spin-$J$ as 
\begin{align}
\phi'^A (x')&= \phi^A (x)- \sigma (x) \vartriangle \phi^A (x) + 
\frac{1}{2}\omega_{\mu\nu}(x) (\Sigma^{\mu\nu})\indices{^A_B} \phi^B (x)\nonumber \\
&=\phi^A (x'-\epsilon)- \sigma (x'-\epsilon) \vartriangle \phi^A (x'-\epsilon) + 
\frac{1}{2}\omega_{\mu\nu}(x'-\epsilon) (\Sigma^{\mu\nu})\indices{^A_B} \phi^B (x'-\epsilon),
\end{align}
where $\vartriangle$ is the specific scale dimension pertaining to the field, and $(\Sigma^{\mu\nu})\indices{^A_B}$ is the spin-$J$ matrix representation. 

Therefore, to first-order approximation, we can deduce the infinitesimal field transformation rule  
\beq
\delta \phi^A (x) \equiv \phi'^A (x) - \phi^A (x)= - \partial_\alpha \phi^A(x) \epsilon ^\alpha (x) 
- \sigma (x) \vartriangle \phi^A (x) + 
\frac{1}{2}\omega_{\mu\nu}(x) {(\Sigma^{\mu\nu})}\indices{^A_B} \phi^B (x).
\label{rule}
\eeq

In particular, for a spin-$1$ vector field $V^\mu(x)$,  
\beq
\vartriangle= \frac{d-2}{2}  \text{\;and\;} (\Sigma^{\mu\nu})\indices{^A_B}=g\indices{^\mu ^A} \delta\indices{^\nu _B} - \delta \indices{^\mu _B} g\indices{^\nu ^A}.
\eeq

Thus, under the SCT, we have
\beq
\delta^\sigma V^\mu(x) = \big( x^2 \partial^\sigma - 2 x^\sigma x\cdot \partial - (d-2) x^\sigma \big) V^\mu(x) + 2x^\mu V^\sigma(x) - 2 g^{\sigma\mu }x\cdot V (x).
\label{V}
\eeq

%\equiv K^\sigma V^\mu(x)It is important to distinguish $K^\sigma $ and $\delta^\sigma$ which will become crucial to our discussion, where the former means transforming a field like a vector under SCT, and the latter is the generic expression of SCT for any field, and should not be restricted to the vector representation.

\section{Is Maxwell Theory Conformally Invariant?}
\subsection{Classical Maxwell Theory}
In this section, we review the literature on the conformal symmetry of classical Maxwell theory.

The free Maxwell theory, with the Lagrangian density
\beq
\Lagr(x)= -\frac{1}{4}F^{\mu\nu}(x)F_{\mu\nu}(x),
\label{10}
\eeq
is known to have scale invariance in any dimension.

To see whether or not the classical theory is also conformally invariant, we will calculate how the Lagrangian transforms under the SCT. 
If the change of $\Lagr(x)$ in (\ref{10})  can be written as a total derivative, then the action would be conformally invariant.

Let us first consider the SCT on the field strength $F^{\mu\nu}(x)=\partial^\mu A^\nu(x) - \partial^\nu A^\mu(x)$. Recall that in a scalar field theory, $\partial_\mu \phi(x)$, the derivative of the primary scalar field, would not transform as a primary vector under the SCT. The SCT on $\partial_\mu \phi(x)$ is defined such that the partial derivative acts only \textit{after} the SCT acts on the primary field $\phi(x)$; hence $\delta^\sigma   \big(\partial_\mu \phi(x)\big)\equiv \partial_\mu \big(\delta^\sigma \phi(x)\big)$. 
Analogously, since $F^{\mu\nu}$ is a descendant of the primary vecotr field $A^\mu$, it would not transform as a primary $2$-form under the SCT. The SCT on $F^{\mu\nu}$ is defined as $\delta^\sigma F^{\mu\nu}(x) \equiv \partial^\mu \delta^\sigma A^\nu(x)-\partial^\nu \delta^\sigma A^\mu(x)$, where again, derivatives acts after the SCT on $A^\mu(x)$. Thus, by using the infinitesimal transformation for $A^\mu (x)$, 
\beq
\delta^\sigma A^\mu(x) = \big( x^2 \partial^\sigma - 2 x^\sigma x\cdot \partial - (d-2) x^\sigma \big) A^\mu(x) + 2x^\mu A^\sigma(x) - 2 g^{\sigma\mu }x\cdot A(x),
\label{A}
\eeq
one deduces the SCT on $F^{\mu\nu}(x)$\cite{Jackiw:2011vz} : 
\begin{align}
\delta^\sigma F^{\mu\nu}(x) =&\; \partial^\mu \delta^\sigma A^\nu(x)-\partial^\nu \delta^\sigma A^\mu(x)\nonumber \\ 
 =&\; (x^2\partial^\sigma - 2 x^\sigma x\cdot \partial - d x^\sigma)F^{\mu\nu}(x) + 2 x^\mu F^{\sigma\nu}(x) - 2 x^\nu F^{\sigma\mu}(x) \nonumber \\ 
 &+ 2g^{\sigma \mu}x_\alpha F^{\nu \alpha}(x)-2g^{\sigma \nu}x_\alpha F^{\mu \alpha}(x) + (d-4)[g^{\sigma\nu} A^\mu(x) - g^{\sigma \mu }A^\nu(x)] 
\label{f}.
\end{align}

The last term shows explicitly that $F^{\mu\nu}$ does not transform as a primary field in $d\neq 4$. Moreover,  the SCT of the physical observable strength $F^{\mu\nu}$ is gauge dependent for $d\neq4$. 

In fact, the gauge non-invariance of (\ref{f}) cannot be compensated by modifying the usual SCT with an additional associated infinitesimal gauge transformation. To see this, let us assume the following modified transformation as a trial:
\begin{equation}
\delta^{'\sigma} A^\mu \equiv \delta^\sigma A^\mu + \partial^\mu \alpha^\sigma,
\label{17}
\end{equation}
which is the combination of the SCT and an infinitesimal gauge transformation. It follows that 
\begin{align}
\delta^{'\sigma} F^{\mu\nu}&= \delta^{\sigma} F^{\mu\nu} + (\partial^\mu \partial^\nu \alpha^\sigma - \partial^\nu \partial^\mu \alpha^\sigma) \nonumber \\
&= \delta^{\sigma} F^{\mu\nu}. 
\end{align}
We see that after the additional gauge transformation, $F^{\mu\nu}$ still remains the same as in (\ref{f}). Thus, it is not possible to remove the gauge dependence in (\ref{f}) by the modified transformation defined in (\ref{17}).  Therefore, classically, the usual SCT is not compatible with gauge symmetry, and it is not a valid symmetry transformation for the gauge theory.

In addition, one finds
\begin{align}
\delta^\sigma \Lagr(x)&= -\frac{1}{2}F_{\mu\nu}(x) \delta^\sigma F^{\mu\nu}(x)\nonumber \\
&=- \partial_\mu \big[(2x^\sigma x^\mu - g^{\sigma \mu} x^2)\Lagr(x)\big] + (d-4) F^{\sigma \mu}(x) A_\mu(x). 
\label{lagr}
\end{align}
The change of the Lagrangian can be expressed as a total derivative only in $d=4$. Furthermore, it is gauge invariant only in $d=4$. Therefore, for $d \neq 4$,  the classical Maxwell theory is not conformal invariant. 

As an aside, using (\ref{f}), one can show the classical equation of motion (EOM)
$\partial_\mu F^{\mu\nu}=0$ is also  not invariant under the SCT in $d\neq 4$ : 
\begin{align}
& \partial_\mu (\delta^\sigma F^{\mu\nu}(x))  \nonumber \\
=& \big(x^2\partial^\sigma - 2x^\sigma x\cdot \partial - (d+2)x^\sigma\big)\partial_\mu F^{\mu\nu}(x)+ 2x^\nu\partial_\mu F^{\mu\sigma}(x) - 2g^{\sigma \nu} x_\alpha \partial_\mu F^{\mu\alpha}(x) \nonumber \\
&\;\; + (d-4)[g^{\sigma\nu} \partial \cdot A(x) - \partial^\nu A^\sigma(x)] \nonumber \\
=& (d-4)[g^{\sigma\nu} \partial \cdot A(x) - \partial^\nu A^\sigma(x)] \nonumber \\ 
\neq& 0.
\label{14}
\end{align}
This result is expected since the action of Maxwell theory is SCT invariant only in $d=4$, and the non-invariance of the EOM is simply a reflection of the non-invariance of the classical theory. 

\subsection{Recovering Conformal Symmetry of Maxwell Theory in the Faddeev-Popov Gauge with $\xi= d/(d-4)$}
Even though the gauge-invariant Maxwell theory does not have the SCT invariance in $d\neq 4$, El-Showk \textit{et al.} \cite{ElShowk:2011gz} argued that the gauge-fixed theory with
\beq
\Lagr_{\xi}(x)= -\frac{1}{4} F^{\mu\nu}(x)F_{\mu\nu}(x) - \frac{1}{2\xi} \big(\partial_\mu A^\mu(x)\big)^2
\label{15}
\eeq
is invariant under the SCT  if $\xi= d/(d-4)$. In their paper, El-Showk \textit{et al.} proved the invariance by showing the quantum two-point correlation function $ \braket{ A^\mu(x) A^\nu(0)}$ does not change under the SCT. 

As an alternative, we will check the SCT invariance of the gauge-fixed theory by explictly computing how the classical action and EOM transform under the SCT. 

The SCT of the gauge-fixed Lagrangian can be calculated by using (\ref{A}) and (\ref{lagr}), 
\begin{align}
\delta^\sigma \Lagr_\xi(x)=- \partial_\mu [(2x^\sigma x^\mu - g^{\sigma \mu} x^2)\Lagr_\xi(x)] + (d-4) F^{\sigma \mu}(x) A_\mu(x) - \frac{d}{\xi} A^\sigma(x) \partial \cdot A(x).
\end{align}

The gauge-fixed action is invariant under the SCT only when $\xi=d/(d-4)$. In this case,  the last two terms can be combined, and the change of the Lagrangian becomes a total derivative 
\begin{align}
\delta^\sigma \Lagr_\xi(x)=- \partial_\mu \big[(2x^\sigma x^\mu - g^{\sigma \mu} x^2)\Lagr_\xi(x)\big] + (d-4)
\partial^\sigma \big(\frac{A^2(x)}{2} \big).
\end{align}
 
Furthermore, the terms in the classical EOM transform under the SCT as:
\begin{align}
& \delta^\sigma \big [\partial_\mu F^{\mu\nu}(x) +\frac{1}{\xi} \partial^\nu \big(\partial \cdot A(x)\big) \big] \nonumber \\
=& \big(x^2\partial^\sigma - 2x^\sigma x\cdot \partial - (d+2)x^\sigma\big)\big[ \partial_\mu F^{\mu\nu}(x)+\frac{1}{\xi} \partial^\nu \big(\partial \cdot A(x)\big) \big] 
+ 2x^\nu\big[\partial_\mu F^{\mu\sigma}(x)) +\frac{1}{\xi} \partial^\sigma \big(\partial \cdot A(x)\big)\big] \nonumber \\ 
&\;\; - 2g^{\sigma \nu} x_\alpha \big[ \partial_\mu F^{\mu\alpha}(x) +\frac{1}{\xi} \partial^\alpha \big(\partial \cdot A(x)\big) \big ]
+ (d-4- \frac{d}{\xi})[g^{\sigma\nu} \partial \cdot A(x) - \partial^\nu A^\sigma(x)]. 
\end{align}
All terms except the last term are proportional to the terms in the original EOM $( \partial_\mu F^{\mu\nu} + \frac{1}{\xi}\partial^\nu \partial\cdot A)$.  Therefore, only when $\xi=d/(d-4)$ does the last term vanish, and the EOM is SCT invariant.

The classical gauge-fixed theory with $\xi=d/(d-4)$ is SCT invariant in any $d\geqslant 3$. This is true in contradiction to the fact that the gauge-invariant Maxwell theory is generally not SCT invariant except in $d=4$. If $d=4$, then $\xi \rightarrow \infty$, the gauge fixing term vanishes, and the gauge invariant theory by itself is SCT invariant, which is consistent with the conclusions of the previous section.

In classical field theory, the gauge-invariant Maxwell theory and the gauge-fixed theory are two distinct theories. In contrast, in quantum field theory, from the path integral point of view, the expectation values of gauge invariant operators should be independent of specific gauge choices. Does this distinction imply that the quantum Maxwell theory is SCT invariant since it is invariant in the particular gauge of $\xi=d/(d-4)$? 

A critical fact noticed by El-Showk \textit{et al.} is that the quantum two-point function $\braket{ A^\mu(x) A^\nu(0)}$ of the gauge-fixed theory defined in (\ref{15}) with $\xi=d/(d-4)$ violates unitarity. Thus, the gauge-fixed theory, contrary to the gauge-invariant Maxwell theory defined in (\ref{10}), is not a unitary theory. However, the two-point function is not gauge invariant nor a physical observable. Thus, it is an intriguing question whether there exists a different gauge-invariant two-point function defined by the gauge-invariant Maxwell theory in (\ref{10})  which is SCT invariant and also satisfies unitarity.

Motivated by the above questions, we shall study in more detail in the next section how the SCT operator acts on the physical/gauge invariant sector of the quantum Maxwell theory.

\subsection{Abelian Quantum Gauge Theory and the BRST Formalism}\label{bb}
In quantum field theory, the compatibility of a symmetry transformation $\mcal O$ with gauge symmetry is not as strict as its classical counterpart -- one need not require $\mcal O$ to be compatible  with gauge symmetry in the entire Hilbert space. As long as the two transformations commute within the physical sector, $\mcal O$ is a symmetry of the quantum theory. 

One common tool for studying the Hilbert space of a quantum gauge theory is the BRST formalism. In the Faddeev-Popov procedure, there is an extra gauge fixing term plus a term involoving ghost $c$ and anti-ghost $\bar c$ degrees of freedom(d.o.f) in addition to the gauge-invariant Lagrangian : 
\beq
\Lagr_{\mathcal {FP}}(x)= -\frac{1}{4} F^{\mu\nu}(x)F_{\mu\nu}(x) - \frac{1}{2\xi} \big(\partial_\mu A^\mu(x)\big)^2 +\partial_\mu  \bar c(x) \partial^\mu c(x).
\eeq
The gauge-fixed Lagrangian, in spite of losing full gauge invariance, is still invariant under a residual symmetry transformation -- the BRST transformation. 

The BRST generator $Q_B$ is nilpotent $(Q_B^2=0)$, and the BRST transformation can be interpreted as a global  gauge transformation within Lorenz gauge $(\partial_\mu A^\mu=0)$ : 
\begin{align}
&[Q_B, A^\mu(x)] = i \partial^\mu c(x) \label{brst1} \\
&\{Q_B, \bar c(x)\} = - i \frac{\partial_\mu A^\mu(x)}{\xi}\\
&\{Q_B, c(x)\} = 0 \label{brst3} . 
\end{align}  
Therefore, we can classify the Hilbert space into three sectors:
\begin{align}
&\mcal H_1 : \ket{\psi_1} \text{\;satisfies\;}  Q_B \ket{\psi_1}\neq 0 \nonumber \\
&\mcal H_2 : \ket{\psi_2}= Q_B \ket{\psi_1} \text{\;and\;}  Q_B \ket{\psi_2} =0 \text{\;(image)} \nonumber \\
&\mcal H_0 : Q_B \ket{\psi_0} =0, \text{\;for which} \ket{\psi_0}\neq Q_B \ket{\psi_1} \text{\;(cohomology)} \nonumber \\
\end{align}
Since the physical states are gauge invariant, they must satisfy $Q_B \ket {phys}=0$. Furthermore, the states in $\mcal H_2$ cannot be physical states since they have zero norm. Therefore, in a theory with BRST symmetry, the only possible physical Hilbert space is $\mcal H_0$, the cohomology of the BRST charge.

The mode expansion for the photon on-shell configuration reads:
\beq
A^\mu(x)= \sum_{\lambda}\int \frac{\dd^{d-1}k}{{(2\pi)}^{d-1}} \frac{1}{\sqrt{2 E_p}} [ \varepsilon^\mu_\lambda (\mathbf k) a^\dagger_\lambda (\mathbf k) e^{ i k\cdot x} + \varepsilon ^{\ast\mu}_\lambda (\mathbf k) a_\lambda (\mathbf k) e^{ -i k\cdot x}],
\eeq
where $a^\dagger_\lambda (\mathbf k)$ and $a_\lambda (\mathbf k)$ are the creation and annihilation operators for the $A^\mu$ field, and $[a_\lambda (\mathbf k),a^\dagger_{\lambda'} (\mathbf k') ]= {(2\pi)}^{d-1} \delta^{(d-1)} (\mathbf k - \mathbf k') \delta_{\lambda \lambda'} $. The symbol $\lambda$ denotes the photon polarizations in $d$ dimension. For example, if $k^\mu= \omega(1,0,0,...,1)$, then 
\begin{align}
\varepsilon^\mu_+(\mathbf k)&= \frac{1}{\sqrt{2}} (1,0,0,...,1)\nonumber\\
\varepsilon^\mu_-(\mathbf k)&= \frac{1}{\sqrt{2}} (1,0,0,...,-1)\nonumber\\
\varepsilon^\mu_{\perp_i}(\mathbf k)&=(0,...,0,i,0,...,0), \; i=1,2,...,d-2,
\end{align}
where $\varepsilon^\mu_+(\mathbf k)$ and $\varepsilon^\mu_-(\mathbf k)$ are the forward and backward polarization vectors respectively, and $\varepsilon^\mu_{\perp_i}(\mathbf k)$ denotes the $d-2$ transverse polarizations.

%the forward polarization $\varepsilon^\mu_+(\mathbf k)= \frac{1}{\sqrt{2}} (1,0,0,...,1)$, the backward polarization  $\varepsilon^\mu_-(\mathbf k)= \frac{1}{\sqrt{2}} (1,0,0,...,-1)$, and the $(d-2)$ transverse polarizations are $\varepsilon^\mu_{\perp_1}(\mathbf k)=(0,1,0,...,0)$, $\epsilon^\mu_{\perp_2}(\mathbf k)=(0,0,1,...,0)$, etc.      

Similarly, the mode expansion of the ghost field is  
\beq
c(x)= \int \frac{\dd^{d-1}k}{{(2\pi)}^{d-1}} \frac{1}{\sqrt{2 E_p}} [  c^\dagger(\mathbf k) e^{ i k\cdot x} + c(\mathbf k) e^{ -i k\cdot x} ] ,
\eeq
with the ghost field creation/annihilation operators  $c^\dagger(\mathbf k)/c(\mathbf k)$, and $\{c(\mathbf k), c^\dagger(\mathbf k')\} = {(2\pi)}^{d-1} \delta^{(d-1)} (\mathbf k - \mathbf k')$.

Applying the above mode expansions to (\ref{brst1}) and (\ref{brst3}), one finds 
\begin{align}
&[Q_B,a^\dagger_\lambda (\mathbf k)]= \sqrt{2w} \delta_{\lambda +}\;\, c^\dagger (\mathbf k) \nonumber \\
&\{Q_B, c^\dagger(\mathbf k)\}=0 .
\end{align}

In particular, for the $d-2$ transverse modes, 
\beq
[Q_B,a^\dagger_{\perp_i} (\mathbf k)]=0, \; i=1,2,...,d-2.
\eeq
Indeed, the transverse modes $a^\dagger_{\perp_i} (\mathbf k)$ in the physical Hilber space $\mcal H_0$ are BRST invariant.

Now, we are ready to investigate the SCT of the physical modes. 
A sensible transformation $\mcal O$ compatible with gauge symmetry should only map one physical state to another physical state. Therefore, if a transformation maps a physical state outside of $\mcal H_0$, then it is not a well-defined transformation. In other words, for Maxwell theory, a physical transformation $\mcal O$ must satisfy:
\beq
[[Q_B,\mcal O], a^\dagger_{\perp_i} (\mathbf k)]=0.
\eeq
That is, the BRST charge must commute with the transformation $\mcal O$ in the physical Hilbert space. It can be readily checked that this condition is satisfied for the Poincare transformations and dilatation. 

Let us now examine whether or not the SCT is compatible with the BRST transformation on the physical modes by computing $[[Q_B,\delta^\sigma], a^\dagger_{\perp_i} (\mathbf k)]$, where $\delta^\sigma$ is the SCT generator and its operation on the gauge field is defined in (\ref{A}).  
We first find for the full gauge field 
\beq
[[Q_B,\delta^\sigma], A^\mu (x)]=-[2x^\sigma \partial^\mu + (d-2) g^{\sigma\mu}]c(x),
\label{commut}
\eeq
where we have used the SCT for the ghost $\delta^\sigma c(x)= [x^2\partial^\sigma- 2x^\sigma x \cdot \partial - (d-2)x^\sigma]c(x)$ since it is a scalar field with scale dimension $\vartriangle = \frac{d-2}{2}$ \cite{ElShowk:2011gz}.

As an example, we take Lorenz gauge and write the mode expansion in a convenient form :
\begin{align}
 A^\mu(x) & \sim \sum_{\lambda}\int \frac{d^{d}k}{{(2\pi)}^{d}} \; [\;\varepsilon^\mu_\lambda (\mathbf k) a^\dagger_\lambda (\mathbf k) e^{ i k\cdot x} + \varepsilon ^{\ast\mu}_\lambda (\mathbf k) a_\lambda (\mathbf k) e^{ -i k\cdot x}\;]\; \delta(k^2) \\
 c(x)  & \sim \int \frac{d^{d}k}{{(2\pi)}^{d}}  \;[\; c^\dagger(\mathbf k) e^{ i k\cdot x} + c(\mathbf k) e^{ -i k\cdot x} \;]\;\delta(k^2).
\end{align}

Combining with (\ref{commut}), we obtain
\beq
[[Q_B,\delta^\sigma], a^\dagger_{\perp_i} (\mathbf k)] \sim (d-4) \varepsilon^\sigma_{\perp_i} c^\dagger (\mathbf k).
\eeq

Therefore, even for the physical modes the SCT is not a symmetry in $d \neq 4$ because the SCT maps the physical modes out of $\mcal H_0$ to the unphysical Hilbert space.  This implies the SCT is a valid transformation only in $d=4$ for quantum Maxwell theory. 

\section{SCT for the Physical and Gauge Dependent Fields }
In the previous sections, we have seen that the SCT is gauge dependent for both classical and quantum Maxwell theories  when $d \neq 4$. In this section, we shall employ a method\footnote{It is helpful to avoid using the mode expansion directly since it is gauge dependent and valid only on shell.} of decomposing the full gauge field into physical and gauge dependent components. This will allow us to compute explicitly how the physical and unphysical d.o.f transform under the SCT and understand how they mix under such a transformation.

\subsection{Gauge Field Decomposition}
The $U(1)$ gauge field $A^\mu (x)$, in both classical and quantum theories, can be decomposed into longitudinal and transverse components  which explicitly preserve the Lorentz invariance of the theory \cite{Leader:2013jra}:  
\begin{align}
& A^\mu(x)=A^\mu_T(x)+ A^\mu_L(x) \\
& A^\mu_L(x) = \frac{1}{\square}\partial^\mu (\partial \cdot A(x)) \label{34}\\
& A^\mu_T(x) = A^\mu(x) - \frac{1}{\square}\partial^\mu (\partial \cdot A(x)) \label{35}, 
\end{align}
where $\square \equiv \partial^\mu \partial_\mu,\;  \mu=0,1,...,d-1$ in $d$ dimension.

By definition, the transverse component is divergence free $(\partial \cdot A_T=0)$, and therefore $\partial \cdot A=\partial \cdot A_L$.  Furthermore, the fact that the longitudinal component does not contribute to the field strength $(\partial^\mu A^\nu_L - \partial^\nu A^\mu_L=0)$ is an indication that $A_L$ corresponds to the unphysical gauge d.o.f. 

Under a gauge transformation 
\beq
A^\mu(x) \rightarrow A'^\mu(x)=A^\mu(x) + \partial^\mu \alpha (x).
\eeq 
Since the transversality condition requires $\partial \cdot A_T'=0$, and in general $\square \alpha \neq 0$,  then the gauge transformation of the transverse and longitudinal components is
\begin{align}
& A^\mu_T(x) \rightarrow A'^\mu_T(x)=A^\mu_T(x)\\
& A^\mu_L(x) \rightarrow A'^\mu_L(x)=A^\mu_L(x)+ \partial^\mu \alpha (x).  
\end{align}
Clearly, $A^\mu_T$ is gauge invariant\footnote{In fact, under a gauge transformation, the most general expression is $A^\mu_T(x) \rightarrow A'^\mu_T(x)=A^\mu_T(x) - \partial^\mu \beta(x)$, and $A^\mu_L(x) \rightarrow A'^\mu_L(x)=A^\mu_L(x)+ \partial^\mu \alpha (x)+ \partial^\mu \beta(x)$, for any $\beta(x)$ satisfying $\square \beta (x)=0$. The term involving $\beta(x)$ corresponds to the residual gauge degree of freedom after fixing the primary gauge. The freedom of choosing $\beta(x)$ without affecting the primary gauge condition is often referred to as the residual Stueckelberg symmetry \cite{Lorce:2012rr}. Once a residual gauge condition $\beta(x)$ is chosen, there is no ambiguity in the definition of $A^\mu_T$. For our purposes, we will set $\beta=0$ throughout this paper.}, \textit{i.e.} $A^\mu_L$ fully accounts for the change of $A^\mu$ under a gauge transformation. One can thus interpret $A^\mu_T$ as the physical field, and $A^\mu_L$ as the unphysical gauge d.o.f in the theory. 

In addition, the gauge field decomposition in (\ref{34}) and (\ref{35}) is indeed consistent with the classification of the BRST Hilbert space since 
\begin{align}
&[Q_B, A^\mu_T(x)] = 0 \label{brst4} \\
&[Q_B, A^\mu_L(x)] = i \partial^\mu c(x),
\end{align}
where we have used (\ref{brst1}).

Therefore, we can attribute $A^\mu_T$ to the physical\footnote{Eq.(\ref{brst4}) only tells us $A^\mu_T$ is the kernel of the BRST charge and belongs to a superposition of $\mcal H_0$ and $\mcal H_2$ Hilbert space. The residual d.o.f from $\mcal H_2$ can be understood as the aforementioned residual Stueckelberg symmetry, where $\beta(x)$ controls how much of $A^\mu_T$ is in $\mcal H_2$. Once we fix $\beta=0$, $A^\mu_T$ can be identified as $\mcal H_0$.}  Hilbert space $\mcal H_0$, and $A^\mu_L$ to the unphysical sector $\mcal H_1$ in the BRST formalism.

\subsection{SCT of $A^\mu_L(x)$ and $A^\mu_T(x)$}
In this section, we will define the SCT on $A^\mu_T$ and $A^\mu_L$. We will show the SCT is not a physical transformation since the transformation of $A^\mu_T$ also depends on $A^\mu_L$, whereas in the case of Poincare transformations and dilatation there is no mixing between the physical and gauge-dependent components. 

By definition, the action of the SCT on $A^\mu_L(x)$ and $A^\mu_T(x)$ is \cite{E.Fradkin1996}
\begin{align}
& \delta^\sigma A^\mu_L(x)\equiv \frac{1}{\square}\partial^\mu \partial_\alpha[\delta^\sigma A^\alpha(x)] \\
& \delta^\sigma A^\mu_T(x)\equiv \delta^\sigma A^\mu(x)-\frac{1}{\square}\partial^\mu \partial_\alpha[\delta^\sigma A^\alpha(x)].
\label{SCTT}
\end{align}
Notice that the decomposition into transverse and longitudinal components acts only \textit{after} transforming the full gauge field under the SCT to ensure that the transversality condition is satisfied, and $\partial_\mu\delta^\sigma A^\mu_T=0$. However, this condition does not guarantee that the SCT of $A^\mu_T$ only depends on $A^\mu_T$ itself.

The task remains is to use the transformation of the primary field in (\ref{A}) to derive the transformation of the transverse and longitudinal components defined above. A detailed calculation can be found in Appendix \ref{sc}. 
The results are
\begin{align}
& \delta^\sigma A^\mu_T(x)= K^\sigma A^\mu_T(x) - d\frac{1}{\square} \partial^\mu A^\sigma_T(x) - (d-4)\frac{1}{\square}[\partial^\mu A^\sigma_L(x) - g^{\sigma \mu} \partial \cdot A_L(x)]
\label{SCTB}\\
& \delta^\sigma A^\mu_L(x)= K^\sigma A^\mu_L(x) + d\frac{1}{\square} \partial^\mu A^\sigma_T(x) + (d-4)\frac{1}{\square}[\partial^\mu A^\sigma_L(x) - g^{\sigma \mu} \partial \cdot A_L(x)], \label{L}
\end{align} 
where $K^\sigma$ denotes the action of the SCT on a vector field :  
\beq
K^\sigma A^\mu_{T(L)}(x) = \big( x^2 \partial^\sigma - 2 x^\sigma x\cdot \partial - (d-2) x^\sigma \big) A^\mu_{T(L)}(x) + 2x^\mu A^\sigma_{T(L)}(x) - 2 g^{\sigma\mu }x\cdot A_{T(L)} (x),
\label{vector}
\eeq
and
\beq
\delta^\sigma A^\mu(x)=K^\sigma A^\mu(x)=K^\sigma A^\mu_T(x)+ K^\sigma A^\mu_L(x).
\label{53}
\eeq

Unlike the full gauge field $A^\mu$ which transforms as a vector under the SCT, the SCT of $A^\mu_T$ and $A^\mu_L$ 
contain non-homogeneous parts in addition to the usual vector transformation. This is due to the fact that the SCT transformation does not commute with the decomposition of the full gauge field into $A^\mu_L$ and $A^\mu_T$. Therefore they cannot be transformed as normal vectors under the SCT. Furthermore, $\partial \cdot A_T$ transforms like a conformal scalar under the SCT:
\beq
\partial_\mu(\delta^\sigma A^\mu_T (x))= (x^2\partial^\sigma - 2 x^\sigma x \cdot \partial -d x^\sigma )\partial \cdot A_T (x). 
\label{461}
\eeq
This is in contrast to $\partial \cdot A$, for which the SCT contains an extra term in addition to the transformation of a primary scalar field:
\beq
\partial_\mu (\delta^\sigma A^\mu (x))= (x^2\partial^\sigma - 2 x^\sigma x \cdot \partial -d x^\sigma )\partial \cdot A (x)+ d A^\sigma(x).
\eeq
Eq.(\ref{461}) provides another check that the transversality condition is ensured after the SCT.

\section{The Extended Conformal Symmetry of Maxwell Theory}  

\subsection{The Extended Conformal Transformation}
Let us further investigate the action of the SCT applied to  $A^\mu_T(x)$ and $A^\mu_L(x)$. At first glance, the transformation of $A^\mu_L$ in (\ref{L}) looks problematic, since it contains a term proportional to $ \partial^\mu A^\sigma_T$. Nonetheless, this term does not affect the physical field strength because $\partial^\mu\partial^\nu A^\sigma_T - \partial^\nu\partial^\mu A^\sigma_T=0$. Thus, one can get rid of the $A^\mu_T$ term in (\ref{L}) by a gauge transformation $A^\mu_L \rightarrow A^{\mu}_L + \partial^\mu \alpha$, with $\alpha= - d\frac{1}{\square} \varepsilon \cdot A_T$.

On the other hand, when one applies the SCT to $A^\mu_T$, the $A^\mu_L$ term in (\ref{SCTB}) mixes the transverse and longitudinal components for $d\neq 4$. Since $A^\mu_T$ is gauge invariant, it is not possible to remove the  $A^\mu_L$ dependence by a gauge transformation. Indeed, the SCT maps the physical field in $\mcal H_0$ to the non-physical $\mcal H_1$ space, and this confirms the result obtained from studying the BRST Hilbert space of this theory in Section 3.3 that the SCT is not a valid transformation the physical Hilbert space in $d\neq 4$. 

Now we will show that it is possible to define an \textit{extended special conformal transformation}(ESCT)\footnote{An alternative definition of the ESCT is given in Appendix \ref{a3}. The alternative ESCT is the combined operation of a finite gauge transformation to Lorenz gauge, followed by the usual SCT and the inverse of the finite gauge transformation. With this definition, $A^\mu_T$ has the same transformation properties as in this section; however the transformation of the unphysical field $A^\mu_L$ is altered. In this alternative approach, the ESCT can be understood as the the gauge invariant extension of the SCT in Lorenz gauge. }, which meets all the requirements of a valid transformation compatible with gauge symmetry : 
\begin{itemize}
\item It only maps one physical state to another, and does not mix unphysical and physical degrees of freedom. 
\item It preserves the transversality condition, $ \partial_\mu \widetilde \delta^\sigma A^\mu_T(x)=0$.
\item It ensures that the longitudinal component is pure gauge and does not contribute to the field strength, \textit{i.e.} $\widetilde \delta^\sigma \big(\partial^\mu A^\nu_L(x) - \partial^\nu A^\mu_L(x) \big)=0$. 
\end{itemize}
Furthermore, we will show that the ESCT is a symmetry for both classical and quantum Maxwell theories. 

In contrast to (\ref{SCTB}) and (\ref{L}), the ESCT on $ A^\mu_L(x)$ and $A^\mu_T(x)$ is defined as :
\begin{align}
\widetilde\delta^\sigma A^\mu_T(x)&\equiv K^\sigma A^\mu_T(x) - d\frac{1}{\square} \partial^\mu A^\sigma_T(x) \label{cor}\\
&=\delta^\sigma A^\mu_T(x) + (d-4)\frac{1}{\square}[\partial^\mu A^\sigma_L(x) - g^{\sigma \mu} \partial \cdot A_L(x)]\\
\widetilde\delta^\sigma A^\mu_L(x)&\equiv K^\sigma A^\mu_L(x) + (d-4)\frac{1}{\square}[\partial^\mu A^\sigma_L(x) - g^{\sigma \mu} \partial \cdot A_L(x)]\label{md}\\
&=\delta^\sigma A^\mu_L(x) - d\frac{1}{\square} \partial^\mu A^\sigma_T(x), 
\label{59}
\end{align}
where $K^\sigma$ is as in (\ref{vector}).

Note that the ESCT is a non-local transformation. In (\ref{cor}), the ESCT on $A^\mu_T$ acts only within the physical Hilbert space; similarly, the ESCT on $A^\mu_L$ in (\ref{md}) depends only on the longitudinal component. The transversality condition is maintained for $A^\mu_T$: $\partial_\mu(\widetilde\delta^\sigma A^\mu_T)=0$, since $ \frac{1}{\square}(\partial^\mu A^\sigma_L - g^{\sigma \mu} \partial \cdot A_L)$ is divergence free. $A^\mu_L$ is guaranteed to be pure gauge because the second term in (\ref{59}) does not change the field strength.

By combining (\ref{cor}) and (\ref{md}), we can define the action of the ESCT on the vector gauge field :
\begin{align}
\widetilde\delta^\sigma A^\mu(x)&\equiv K^\sigma A^\mu(x) - d\frac{1}{\square} \partial^\mu A^\sigma_T(x) + (d-4)\frac{1}{\square}[\partial^\mu A^\sigma_L(x) - g^{\sigma \mu} \partial \cdot A_L(x)].
\label{full}
\end{align}

It can be verified that the field strength transforms under the ESCT  as 
\begin{align}
\widetilde\delta^\sigma F^{\mu\nu}(x) 
 =&\; (x^2\partial^\sigma - 2 x^\sigma x\cdot \partial - d x^\sigma)F^{\mu\nu}(x) + 2 x^\mu F^{\sigma\nu}(x) - 2 x^\nu F^{\sigma\mu}(x) \nonumber \\ 
 &+ 2g^{\sigma \mu}x_\alpha F^{\nu \alpha}(x)-2g^{\sigma \nu}x_\alpha F^{\mu \alpha}(x) + (d-4)[g^{\sigma\nu} A_T^\mu(x) - g^{\sigma \mu }A_T^\nu(x)].
 \label{51}
\end{align}

Although $F^{\mu\nu}$ still does not transform as a primary tensor field under the ESCT due to the last term in (\ref{51}), $\widetilde \delta^\sigma F^{\mu\nu}$ is independent of gauge choices, in contrast to $\delta^\sigma F^{\mu\nu}$ in (\ref{f}); this is because $F^{\mu\nu}$ is a descendant of the gauge invariant $A^\mu_T$, and $\widetilde\delta^\sigma A^\mu_T$ depends only on the physical field $A_T$ as in (\ref{cor}). Thus, the ESCT is a valid transformation for Abelian gauge theory. The algebra of the extended conformal transformations is discussed in Appendix \ref{algebra} .

\subsection{Invariance of the Maxwell Theory}
In this section, we demonstrate the invariance of the Maxwell action and gauge-invariant correlation functions under the ESCT. 

Using the new transformation, 
\begin{align}
\widetilde\delta^\sigma \Lagr(x)&= -\frac{1}{2}F_{\mu\nu}(x) \widetilde\delta^\sigma F^{\mu\nu}(x)\nonumber \\
&=- \partial_\mu \big[(2x^\sigma x^\mu - g^{\sigma \mu} x^2)\Lagr(x)\big] + (d-4) F\indices{^\sigma_\mu}(x) A^\mu_T(x). 
\label{45}
\end{align}
Compared with $\delta^\sigma \Lagr$ in (\ref{lagr}), the two expressions are almost identical except that  the full gauge field $A^\mu$ is replaced by the physical transverse component $A^\mu_T$ in (\ref{45}). Yet, this tiny modification makes a huge difference. In (\ref{45}), the first term is a trivial total derivative. In fact, the second term  
\begin{align}
F\indices{^\sigma_\mu}A^\mu_T &=(\partial^\sigma A^\mu_T - \partial^\mu A^\sigma_T)A^\mu_T \nonumber \\
&= \partial^\sigma (\frac{A_T^2}{2}) - \partial^\mu(A^\sigma_T A_{T\mu}) + A_T^\sigma (\partial \cdot A_T), 
\label{46}
\end{align}
is actually also a total derivative as guaranteed by the transversality condition $\partial \cdot A_T=0$.
We then have,
\begin{align}
\widetilde\delta^\sigma \Lagr(x) &=- \partial_\mu \big[(2x^\sigma x^\mu - g^{\sigma \mu} x^2)\Lagr(x)\big] + \partial_\mu \Big[(d-4)\big( \frac{g^{\sigma\mu} A^2_T(x)}{2} - A^\sigma_T(x) A^\mu_T(x) \big)\Big].
\label{47}
\end{align}
Indeed, $\widetilde\delta^\sigma \Lagr(x)$ is a total derivative, and therefore the classical Maxwell action would be invariant under the ESCT. 
\\

To study the invariance of the quantum version of the Maxwell theory under the extended SCT, let us examine how the gauge-invariant transverse two-point function $\braket{A^\mu_T(x)A^\nu_T(y)}$ changes under the ESCT.

In momentum-space, the transverse two-point function is defined by 
\beq
\braket{A^\mu_T(k)A^\nu_T(-k)} \equiv \; \frac{1}{k^2}\big(\;  g^{\mu\nu} - \frac{k^\mu k^\nu}{k^2} \; \big). 
\eeq
Its Fourier transformation yields the position-space expression of the transverse two-point function up to an overall normalization constant:
\beq
\braket{A^\mu_T(x)A^\nu_T(0)} = \frac{1}{{(x ^2)}^\vartriangle} \; \big(\; g^{\mu\nu} + 2\vartriangle \frac{x^\mu x^\nu}{x^2} \; \big), 
\label{tt}
\eeq
where $\vartriangle = \frac{d-2}{2}$ is the scale dimension of $A^\mu$. 

Recall that the transverse two-point function is invariant under the Poincare and scale transformations, which are symmetries of
the quantum Maxwell theory. Let us now compute how the transverse two-point function transforms under the ESCT:  
\begin{align}
\widetilde \delta^\sigma \braket{A^\mu_T(x)A^\nu_T(0)} &= \braket{\big(\widetilde\delta^\sigma A^\mu_T(x)\big)A^\nu_T(0)} + \braket{ A^\mu_T(x)\big(\widetilde\delta^\sigma A^\nu_T(0)\big)} \nonumber \\
&= K^\sigma \braket{A^\mu_T(x)A^\nu_T(0)} - d \frac{1}{\square} \big( \; \partial^\mu  \braket{A^\sigma_T(x)A^\nu_T(0)} - \partial^\nu  \braket{A^\mu_T(x)A^\sigma_T(0)} \;\big),
\label{1}
\end{align}
where we have used the ESCT of $A^\mu_T(x)$ in (\ref{cor}).  

Note also that $K^\sigma \braket{A^\mu_T(x)A^\nu_T(0)}$ transforms the two-point function as if its arguments are normal vector fields as defined in (\ref{vector}). Thus, in order to evaluate $K^\sigma \braket{A^\mu_T(x)A^\nu_T(0)}$, let us consider a theory of a vector field $V^\mu$, which is invariant under Poincare and scale transformations. The most general form of its correlation function is
\beq
\braket{V^\mu(x)V^\nu(0)} = \frac{1}{{(x ^2)}^\vartriangle} \; \big(\; c_1 \; g^{\mu\nu} + c_2 \; \frac{x^\mu x^\nu}{x^2} \; \big), 
\eeq
with arbitrary coeffients $c_1$ and $c_2$. 
A detailed calculation gives the SCT of this two-point function :   
\begin{align}
 \delta^\sigma \braket{V^\mu(x)V^\nu(0)}&=K^\sigma \braket{V^\mu(x)V^\nu(0)} \nonumber \\
&= \frac{( 2c_1+ c_2 ) }{{(x^2)}^{\vartriangle}} \big(\; g^{\sigma\nu}x^\mu - g^{\sigma\mu}x^\nu \; \big).
\label{62}
\end{align}
Applying (\ref{62}) to $\braket{A^\mu_T(x)A^\nu_T(0)}$ in (\ref{tt}), we obtain $c_1=1$ and $c_2= 2\vartriangle$, and
\begin{align}
 K^\sigma \braket{A^\mu_T(x)A^\nu_T(0)} = \frac{d }{{(x^2)}^{\vartriangle}} \big(\; g^{\sigma\nu}x^\mu - g^{\sigma\mu}x^\nu \; \big).
 \label{11}
\end{align}

The last term in (\ref{1}) can be simplified  as  
\begin{align}
& -d\frac{1}{\square} \big( \; \partial^\mu  \braket{A^\sigma_T(x)A^\nu_T(0)} - \partial^\nu  \braket{A^\mu_T(x)A^\sigma_T(0)} \;\big) \nonumber \\
= &-d \frac{1}{\square} \Big [\frac{-4 \vartriangle}{{(x^2)}^{\vartriangle +1}} \big(\; g^{\sigma\nu}x^\mu - g^{\sigma\mu}x^\nu \; \big)\Big] \nonumber \\
= & -d \frac{1 }{{(x^2)}^{\vartriangle}} \big(\; g^{\sigma\nu}x^\mu - g^{\sigma\mu}x^\nu \; \big).
\label{22}
\end{align}

Finally, substituting (\ref{11}) and (\ref{22}) into (\ref{1}), we find 
\begin{align}
\widetilde \delta^\sigma \braket{A^\mu_T(x)A^\nu_T(0)}&= \frac{d }{{(x^2)}^{\vartriangle}} \big(\; g^{\sigma\nu}x^\mu - g^{\sigma\mu}x^\nu \; \big) - \frac{d }{{(x^2)}^{\vartriangle}} \big(\; g^{\sigma\nu}x^\mu - g^{\sigma\mu}x^\nu \; \big)\nonumber\\
&=0.
\end{align}
This proves the invariance of the transverse two-point function under the  ESCT.

Since the $U(1)$ gauge theory is a non-interacting free quantum field theory, the proof of the invariance of the transverse two-point function is sufficient for showing that all gauge invariant $n$-point functions of the quantum theory would also be invariant under the ESCT. 

\subsection{Invariance of the Classical Equation of Motion}
A symmetry of the classical action implies the invariance of the classical EOM under the symmetry transformation. Hence, it is necessary to check that the EOM is invariant under the ESCT. 
Although the gauge field decomposition defined in (\ref{34}) and (\ref{35}) is sufficient for calculating off-shell quantities such as the correlation function, it is problematic on shell. For instance, in the Lorenz gauge $\partial_\mu A^\mu=0$, when the EOM $\square A^\mu=0$ is imposed, $A^\mu_L= \frac{1}{\square}\partial^\mu (\partial \cdot A)$ is not well-defined. Thus, to avoid a singularity of the gauge field decomposition when calculating on-shell quantities such as the EOM, we will introduce a source current as a regulator.

The Lagrangian with a source term reads 
\beq
\Lagr_J(x)= -\frac{1}{4}F^{\mu\nu}(x)F_{\mu\nu}(x) +  J^\mu(x) A_\mu(x),
\label{66}
\eeq
where 
\beq
\partial_\mu J^\mu=0,
\label{j}
\eeq
due to conservation of the $U(1)$ charge.
The EOM is 
\beq
\partial_\mu F^{\mu\nu} - J^\nu =0,
\label{h}
\eeq
and by setting $J^\nu=0$, the EOM of the free theory is recovered.

In order to calculate how (\ref{h}) transforms under the ESCT, we need to know the ESCT on $J^\mu$, which is not yet specified. The transformation on $J^\mu$ should be defined so that the change of $J \cdot A$ is a total derivative and then it will not affect the symmetry property of the theory. For example, since the free Maxwell theory is Lorentz invariant, one would expect that the Lorentz symmetry will still be a property of (\ref{66}). Therefore, under a Lorentz transformation, $J^\mu$ must transform as a Lorentz vector so that 
\beq
\delta( J \cdot A) \rightarrow \delta( J \cdot A) - \partial_\mu\big[ (J \cdot A) \omega_{\mu\nu} x^\nu\big],
\eeq
where the change is a total derivative.  

In the case of Poincare and scale transformations, the source current $J^\mu$ indeed transforms as a normal vector with scale dimension $\vartriangle = \frac{d+2}{2}$, and $\partial_\mu J^\mu=0$ is preserved. Conservation of the gauge current under these transformations implies that Poincare and scaling symmetries are properties of the gauge-invariant Maxwell theory. However, for the SCT, it is not obvious whether $J^\mu$ transforms as a primary conformal vector field. In fact, as we shall see,  current conservation will be violated for $d\neq 4$ if $J^\mu$ is treated as a conformal vector in the SCT. Nevertheless, we will find that the gauge current is conserved under the ESCT in any dimension. 
  
Let us first assume that $J^\mu$ transform as a primary vector with $\vartriangle = \frac{d+2}{2}$ under the SCT:
\begin{align}
\delta^\sigma J^\mu(x) = \big( x^2 \partial^\sigma - 2 x^\sigma x\cdot \partial - (d+2) x^\sigma \big) J^\mu(x) + 2x^\mu J^\sigma(x) - 2 g^{\sigma\mu }x\cdot J (x).
\end{align}
It follows that the $J \cdot A$ term in the Lagrangian will transform as a conformal scalar, and $\delta^\sigma (J \cdot A)= - \partial_\mu\big[ (J \cdot A) \delta^\sigma x^\mu \big]$ .

Nevertheless, the SCT of (\ref{j}) then gives:
\begin{align}
\partial_\mu \delta^\sigma J^\mu(x) &=  \big( x^2 \partial^\sigma - 2 x^\sigma x \cdot \partial  - (d+4)x^\sigma \big) \partial \cdot J(x) + (d-4) J^\sigma (x) \nonumber\\
&= (d-4) J^\sigma (x) \neq 0 \;!
\end{align}
We see that after the SCT, $\partial_\mu J^\mu=0$ 
is still true only in $d=4$. For $d\neq 4$, the SCT explicitly breaks gauge invariance of the theory and the $U(1)$ gauge current is not conserved. This fact reveals that the SCT is compatible with gauge symmetry only in $d=4$.

In fact, we can construct an ESCT for $J^\mu$ which preserves current conservation in any dimension.

We will define 
\beq
\widetilde \delta^\sigma J^\mu(x)= \delta^\sigma J^\mu(x) - (d-4)\frac{1}{\square}\partial^\mu J^\sigma(x).
\label{msct}
\eeq 
In this extended transformation, $\partial \cdot J$ transforms as a conformal scalar and the $U(1)$ current is conserved as desired:
\beq
\partial_\mu \widetilde \delta^\sigma J^\mu(x) =  \big[ x^2 \partial^\sigma - 2 x^\sigma x \cdot \partial  - (d+4)x^\sigma \big] \partial \cdot J(x).
\eeq
It can be verified that the ESCT of the $J \cdot A$ term in the Lagrangian is a total derivative which does not affect the symmetry of the theory.

We can now compute how the terms in the EOM transform under the ESCT, given the transformations on $F^{\mu\nu}$ and $J^\mu$ defined in (\ref{51}) and (\ref{msct}) respectively:
\begin{align}
& \widetilde \delta^\sigma (\partial_\mu F^{\mu\nu} - J^\nu) \nonumber \\
=& \big(x^2\partial^\sigma - 2x^\sigma x\cdot \partial - (d+2)x^\sigma\big)(\partial_\mu F^{\mu\nu} - J^\nu)+ 2x^\nu (\partial_\mu F^{\mu\sigma} - J^\sigma) - 2g^{\sigma \nu} x_\alpha (\partial_\mu F^{\mu\alpha} - J^\alpha) \nonumber \\
&\;\; - (d-4) \frac{1}{\square}\partial^\nu \big[\square A^\sigma_T - \partial^\sigma (\partial \cdot A_T) - J^\sigma\big] \nonumber \\
=&  \big(x^2\partial^\sigma - 2x^\sigma x\cdot \partial - (d+2)x^\sigma\big)(\partial_\mu F^{\mu\nu} - J^\nu)+ 2x^\nu (\partial_\mu F^{\mu\sigma} - J^\sigma) - 2g^{\sigma \nu} x_\alpha (\partial_\mu F^{\mu\alpha} - J^\alpha) \nonumber \\
&\;\;  - (d-4) \frac{1}{\square}\partial^\nu (\partial_\mu F^{\mu\sigma} - J^\sigma) \nonumber \\
=& 0. 
\label{73}
\end{align}
To derive the second equality, we have used $\square A^\sigma_T= \partial_\mu F^{\mu\nu}$ and $\partial \cdot A_T=0$ for the last term in the first equality.
Since every term in (\ref{73}) is proportional to the terms in the original EOM, the EOM is thus ESCT invariant. 
Notice that in the free theory, the inverse of the d'Alembert operator on  $A^\sigma_T$ is ill-defined because EOM implies $\square A^\sigma_T=0$. However, after the source current $J^\mu$ is added, one can re-write $A^\sigma_T= \frac{1}{\square} (\square A^\sigma_T)$ without worrying about the on-shell singularity. Thus, this calculation has demonstrated the invariance of the classical EOM under the ESCT. 

In conclusion, the invariance of the classical action together with the invariance of the EOM proves  the invariance of classical Maxwell theory under the ESCT.

\section{Discussion}
In this paper we have discovered a gauge-invariant non-local \textit{extended special conformal transformation}(ESCT), in which both the quantum and classical Maxwell theories are invariant for any dimension $d\geqslant3$ . The ESCT defines a new symmetry  of the gauge invariant physical sector of the Maxwell theory in general dimensions. Because of its enhanced properties, physical states are mapped only within the physical Hilbert space, the field strength $F^{\mu\nu}$ transforms as a descendant of the gauge-invariant $A^\mu_T$,  and the $U(1)$ gauge current is preserved. This is in contrast to the usual SCT, where the gauge invariance of theory is broken because of mixing between the physical and unphysical sectors in the BRST Hilbert space. 

The ESCT provides a new way of understanding the conformal symmetry of gauge theories. Since it is not necessary to require conformal invariance in the unphysical sector of a gauge theory, the ESCT invariance of the physical sector will ensure that all gauge-invariant physical correlation functions are conformally invariant. 

The insights gained from the ESCT can be generalized and applied to the analysis of other scale invariant gauge theories which are normally not considered to be conformally invariant in higher dimensions. 
It will be interesting to investigate the extension of the ESCT to classical non-Abelian gauge theories. This is a subject for future studies.

\section*{Acknowledgements}
The authors have benefited greatly from discussions with Cedric Lorce on the decomposition of gauge field. We are grateful to Michael Peskin, Jose Cembranos, and Alfred Goldhaber for valuable comments and suggestions. YJC would also like to thank Kuang-Ting Chen for useful discussions. This work was supported by the US Department of Energy under contract DE--AC02--76SF00515.

\begin{appendices}
\section{SCT of $A_T$ and $A_L$} \label{sc}
In this appendix, we explicit derive the SCT on $A_T(x)$ and $A_L(x)$ first introduced in (\ref{SCTB}) and (\ref{L}).

Let us first compute the SCT of $A_L(x)$.
\begin{align}
 &\delta^\sigma  A^\mu_L(x) \nonumber\\
 =& \frac{1}{\square}\partial^\mu \partial_\alpha\big[\delta^\sigma A^\alpha(x)\big] \nonumber\\
 =&\frac{1}{\square}\partial^\mu \big[(x^2- 2x^\sigma x\cdot \partial - d x^\sigma)\partial \cdot A(x) + dA^\sigma(x)\big] \label{81}\\
=& \frac{1}{\square}\big[(2x^\mu \partial^\sigma - 2 g^{\sigma \mu} x\cdot \partial - 2 x^\sigma \partial^\mu - d g^{\sigma \mu}) \partial \cdot A(x)\big]   \nonumber\\
&+ \frac{1}{\square}\Big[(x^2- 2x^\sigma x\cdot \partial - d  x^\sigma)\partial^\mu \big(\partial \cdot A(x)\big)\Big] + d \frac{1}{\square} \partial^\mu A^\sigma(x). 
\end{align}

Using the identity\cite{E.Fradkin1996}
\begin{align}
&\square\big[(2x^\mu \partial^\sigma - 2 g^{\sigma \mu} x\cdot \partial - 2 x^\sigma \partial^\mu )\frac{1}{\square} \partial \cdot A(x)\big] \nonumber \\
=& (2x^\mu \partial^\sigma - 2 g^{\sigma \mu} x\cdot \partial - 2 x^\sigma \partial^\mu - d g^{\sigma \mu}) \partial \cdot A(x) +(d-4) g^{\sigma \mu} \partial \cdot A(x),
\end{align}
we can rewrite
\begin{align}
& \frac{1}{\square} \big[(2x^\mu \partial^\sigma - 2 g^{\sigma \mu} x\cdot \partial - 2 x^\sigma \partial^\mu - d g^{\sigma \mu}) \partial \cdot A(x)\big]\nonumber \\
=& (2x^\mu \partial^\sigma - 2 g^{\sigma \mu} x\cdot \partial - 2 x^\sigma \partial^\mu )\frac{1}{\square} \partial \cdot A(x) - (d-4)g^{\sigma \mu}\frac{1}{\square} \partial \cdot A(x).
\end{align}

Similarly,
\begin{align}
& \square\big[(x^2\partial^\sigma - 2 x^\sigma x\cdot \partial) \frac{1}{\square}\partial^\mu \big(\partial \cdot A(x)\big)\big] \nonumber\\
=& (x^2\partial^\sigma - 2x^\sigma x\cdot\partial - d x^\sigma) \partial^\mu \big(\partial \cdot A(x)\big) \nonumber\\
&+ (d-4) x^\sigma \partial^\mu \big(\partial \cdot A(x)\big) + (2d-4)\frac{1}{\square} \partial^\sigma \partial^\mu \big(\partial \cdot A(x)\big),
\end{align}
we rewrite 
\begin{align}
&\frac{1}{\square}\Big[(x^2- 2x^\sigma x\cdot \partial - d  x^\sigma)\partial^\mu \big(\partial \cdot A(x)\big)\Big]\nonumber\\
=& (x^2\partial^\sigma- 2x^\sigma x\cdot \partial)\frac{1}{\square}\partial^\mu \big(\partial \cdot A(x)\big)\nonumber\\
& - (d-4)\frac{1}{\square}\Big[x^\sigma\partial^\mu \big(\partial \cdot A(x)\big)\Big] -(2d-4)\frac{1}{\square^2}\partial^\sigma \partial^\mu \big(\partial \cdot A(x)\big)
\end{align}

Combining all the terms and plugging in  $\displaystyle A^\mu_L(x)= \frac{1}{\square}\partial^\mu \big(\partial \cdot A(x)\big)$ and
\beq
\displaystyle \frac{1}{\square}\Big[x^\sigma\partial^\mu \big(\partial \cdot A(x)\big)\Big]= x^\sigma A^\mu_L (x) - 2 \frac{1}{\square} \partial^\sigma A^\mu_L(x),
\eeq
we find
\beq
\delta^\sigma A^\mu_L(x)= K^\sigma A^\mu_L(x) + d\frac{1}{\square} \partial^\mu A^\sigma_T(x) + (d-4)\frac{1}{\square}[\partial^\mu A^\sigma_L(x) - g^{\sigma \mu} \partial \cdot A_L(x)].
\eeq

Since, $\delta^\sigma A^\mu(x)= \delta^\sigma A^\mu_L(x)+ \delta^\sigma A^\mu_T(x)= K^\sigma A^\mu(x)$, the transformation of $A^\mu_T(x)$ can thus be deduced : 
\beq
 \delta^\sigma A^\mu_T(x)= K^\sigma A^\mu_T(x) - d\frac{1}{\square} \partial^\mu A^\sigma_T(x) - (d-4)\frac{1}{\square}[\partial^\mu A^\sigma_L(x) - g^{\sigma \mu} \partial \cdot A_L(x)].
\eeq

\section{Extended Conformal Algebra}\label{algebra}
The extended conformal algebra is defined  by the commutation relations among the ESCT generators $\widetilde {\mcal K}^\alpha$ and with the other normal conformal generators $\mcal P^\alpha$(translation), $\mcal  M^{\alpha \beta}$(Lorentz rotation), and $\mcal D$(dilatation). 

The original conformal generators form a closed group and have the following commutation relations:
\begin{align}
&[\mcal D, \mcal P^\mu] = i\;\mcal P^\mu  \\
&[\mcal D, \mcal K^\mu] = -i\;\mcal K^\mu \\
&[\mcal K^\mu, \mcal P^\nu] = 2i\left( g^{\mu\nu} \mcal D - \mcal M^{\mu\nu} \right)\\
&[\mcal K^\alpha, \mcal M^{\mu\nu}] = i\left( g^{\alpha\mu} \mcal K^\nu - g^{\alpha\nu} \mcal K^\mu  \right)\\
&[\mcal P^\alpha, \mcal M^{\mu\nu}] = i\left( g^{\alpha\mu} \mcal P^\nu - g^{\alpha\nu} \mcal P^\mu  \right)\\
&[\mcal M^{\alpha\beta}, \mcal M^{\mu\nu}] = i\left( g^{\alpha\nu} \mcal M^{\beta\mu} + g^{\beta\mu} \mcal M^{\alpha\nu} - g^{\alpha\mu} \mcal M^{\beta\nu} - g^{\beta\nu} \mcal M^{\alpha\mu}\right).
\end{align}  

One way to obtain the commutators is to let the generators act on fields and then swap the order. For example, $[{\mcal P}^{\sigma}, \mcal M^{\mu\nu}]$ can be deduced by computing
\begin{align}
[\;[{\mcal P}^{\sigma}, \mcal M^{\mu\nu}], A^\alpha(x)\;] &= [\;{\mcal P}^{\sigma}, [\mcal M^{\mu\nu}, A^\alpha(x)]\;] - [\;\mcal M^{\mu\nu}, [{\mcal P}^{\sigma}, A^\alpha(x)]\;].
\end{align}
The action of conformal generators on the primary vector field $A^\alpha(x)$ is given by 
\begin{align}
[\mcal P^\mu,A^\alpha(x) ] &= -i\,\partial^\mu A^\alpha(x) \label{191} \\ 
[\mcal M^{\mu\nu}, A^\alpha(x)] &= i\,\Big[ \big( x^\mu \partial^\nu - x^\nu\partial^\mu \big)A^\alpha(x) + g^{\alpha\mu}A^\nu(x)- g^{\alpha\nu}A^\mu(x) \Big]\label{192}\\
[\mcal D, A^\alpha(x) ] &= -i\,\big( x^\mu \partial_\mu + \frac{d-2}{2} \big)A^\alpha(x)   \\
[\mcal K^{\mu}, A^\alpha(x)] &= i\,\big[ \big( x^2\partial^\mu - 2x^\mu x\cdot \partial - (d-2)x^\mu\big)A^\alpha(x) + 2x^\alpha A^\mu(x) - 2 g^{\mu\alpha} x\cdot A(x) \big]\nonumber \\[0.1 em]
&\equiv i\,K^\mu A^\alpha(x).
\end{align} 

We shall now compute the extended conformal algebra. Since there is no mixing between physical and unphysical components in the extended conformal transformations, it is thus sufficient to calculate the commutation relations by studying the action of extended conformal generators on the  physical sector spanned by $A^\mu_T$, as defined in (\ref{35}). 

Recall from (\ref{SCTB}) that under the usual SCT,  $A^\alpha_T$ transforms as 
\begin{align}
[\mcal K^{\mu}, A^\alpha_T(x)] &= i\,\big[K^\mu A^\alpha_T(x) - d\frac{1}{\square} \partial^\alpha A^\mu_T(x) - (d-4)\frac{1}{\square}\big(\partial^\alpha A^\mu_L(x) - g^{ \mu \alpha} \partial \cdot A_L(x)\big) \big]\nonumber \\[0.1 em]
&\equiv i\,\delta^\mu A^\alpha_T(x),
\label{101}
\end{align}
and the ESCT on  $A^\alpha_T$ is constructed by removing the  terms involving $A^\alpha_L$ in (\ref{101}). The ESCTs generator thus satisfy 
\begin{align}
[\widetilde{\mcal K}^{\mu}, A^\alpha_T(x)] = i\,\big[ K^\mu A^\alpha_T(x) - d\frac{1}{\square} \partial^\alpha A^\mu_T(x)  \big]\equiv i\,\widetilde \delta^\mu A^\alpha_T(x).
\end{align}

An explicit calculation then gives the modified commutators acting on $A^\alpha_T$
\begin{align}
&[\;[\mcal D, \widetilde{\mcal K}^{\mu}], A^\alpha_T(x)\;] = -i\;[\;\widetilde{\mcal K}^\mu, A^\alpha_T(x)\;] \label{1103}\\
&[\;[\widetilde{\mcal K}^{\mu}, \mcal P^\nu], A^\alpha_T(x)\;] =2i\,[\;  g^{\mu\nu} \mcal D - \mcal M^{\mu\nu} , A^\alpha_T(x)\;]\label{1104}\\
&[\;[\widetilde{\mcal K}^{\sigma}, \mcal M^{\mu\nu}], A^\alpha_T(x)\;] = i\,[\; g^{\sigma\mu} \widetilde{\mcal K}^\nu - g^{\sigma\nu} \widetilde{\mcal K}^\mu, A^\alpha_T(x)\;] \label{1105}\\
&[\;[\widetilde{\mcal K}^{\mu}, \widetilde{\mcal K}^{\nu}], A^\alpha_T(x)\;] = - d(d-4) \frac{1}{\square}\big[\frac{1}{\square} \partial^\alpha F^{\mu\nu}(x) + g^{\alpha\nu}A^\mu_T(x) - g^{\alpha\mu}A^\nu_T(x) \big] \label{106}.
\end{align}  

An intuitive way to understand the results in (\ref{1103})-(\ref{106}) is the following:
Define $P$ to be the projection operator which eliminates the $A^\alpha_L$ terms in $\delta^\sigma A^\alpha_T$, and
\beq
[\widetilde{\mcal K}^{\mu}, A^\alpha_T(x)] = [P{\mcal K}^{\mu}P, A^\alpha_T(x)].
\eeq
Since $\mcal D$,  $\mcal P^\mu$, and  $\mcal M^{\mu\nu}$ act on $A^\alpha_T$ only within the physical sector, they all commute with $P$.
We then have  for example, 
\begin{align}
&[\;[\mcal D, \widetilde{\mcal K}^{\mu}], A^\alpha_T(x)\;]\nonumber\\
=&[\;[\mcal D, P{\mcal K}^{\mu}P], A^\alpha_T(x)\;] \nonumber\\
=& [\;P[\mcal D, {\mcal K}^{\mu}]P, A^\alpha_T(x)\;]\nonumber\\
=&-i\;[\;\widetilde{\mcal K}^\mu, A^\alpha_T(x)\;].
\end{align}

Therefore, the commutation relations in (\ref{1103})-(\ref{1105}) between $\widetilde{\mcal K}^{\mu} $ and other normal conformal generators retain the same form. However, since the ESCT generators do not commute with $P$, they actually do not commute with each other in contrast to the usual SCT generators in $d\neq 4$. 

In addition, it is not possible to express the right hand side(RHS) of (\ref{106}) in terms of the extended conformal generators. One may wonder if this can be remedied by a gauge transformation. The answer is negative, because gauge transformations only act on the gauge dependent component $A^\mu_L$ and does not affect $A^\mu_T$. Also, if one includes the new operators on the RHS  of (\ref{106}) into the extended conformal group as generators and compute their commutation relations, more powers of $\frac{1}{\square}$ would occur. This procedure thus generates an infinite number of new generators. Hence, the extended conformal generators which include $\widetilde {\mcal K}^\alpha$ do not form a closed group.

\section{ Alternative Definition of the ESCT}\label{a3}
In this appendix, we introduce an alternative definition of the ESCT, which has very similar properties to the one defined by (\ref{51}), and it will give a physical understanding of the nature of the ESCT.

Recall that the Maxwell action is SCT invariant even in $d\neq 4$ if $F\indices{^\sigma_\mu}A^\mu $ in (\ref{lagr}) is a total derivative. Since 
\begin{align}
F\indices{^\sigma_\mu}A^\mu &=(\partial^\sigma A^\mu - \partial^\mu A^\sigma)A^\mu \nonumber \\
&= \partial^\sigma (\frac{A^\mu A_{\mu}}{2}) - \partial^\mu(A^\sigma A_{\mu}) + A^\sigma (\partial \cdot A), 
\end{align}
this term is indeed a total derivative if one works in the Lorenz gauge ($\partial \cdot A=0$). That is, given any gauge configuration $A^\mu$, as long as it is transformed to the Lorenz gauge before the SCT is applied, then the action and the classical EOM will be SCT invariant. In fact, one can define a gauge transformation 
\beq
\mcal G A^\mu \equiv A^\mu + \partial^\mu \alpha,
\label{91}
\eeq
which transforms $A^\mu$ to the Lorenz gauge where $A^\mu= A^\mu_T$ and $A^\mu_L=0$. The specific gauge transformation $\mcal G_T$ satisfies 
\beq
\mcal G_T A^\mu = A^\mu_T, 
\label{92}
\eeq
and thus 
\beq
\alpha= -\frac{1}{\square } \partial \cdot A,
\label{93}
\eeq
is a \textit{finite} transformation independent of the \textit{infinitesimal} SCT.

Since the SCT is an  infinitesimal transformation of $\mcal O (\varepsilon^1)$ and $\alpha$ is independent of  $\varepsilon^\mu$ at $\mcal O (\varepsilon^0)$, an additional inverse gauge transformation 
\beq
\mcal G_T^{-1} A^\mu \equiv A^\mu - \partial^\mu  \alpha, \text{\; with\;\;} \alpha= -\frac{1}{\square } \partial \cdot A, \text{ and\;\;} \mcal G_T^{-1}\mcal G_T = \mathbb{1},
\eeq
must be performed after the SCT to make the combined transformation only at $\mcal O (\varepsilon^1)$. Therefore, we can define an \textit{alternative extended special conformal transformation}(aESCT) acting on $A^\mu$ as  
\begin{align}
\dt{\delta^\sigma} 
A^\mu &\equiv \mcal G_T^{-1} \,(1+\delta^\sigma) \; \mcal G_T A^\mu - A^\mu \label{aesct} \\
&= \mcal G_T^{-1} (1+K^\sigma) A^\mu_T - A^\mu \nonumber  \\ 
&= (1+K^\sigma) A^\mu_T - \partial^\mu \alpha - A^\mu \nonumber\\
&= K^\sigma A^\mu_T.
\label{94}
\end{align}
In the second line,  since $\mcal G_T A^\mu$ is a vector field, we can use (\ref{53}) and (\ref{92}) to rewrite $\delta^\sigma \mcal G_T A^\mu =K^\sigma \mcal G_T A^\mu = K^\sigma  A^\mu_T $ . To yield the third and fourth lines, the gauge transformation defined by (\ref{91}) and (\ref{93}) is used. 

Next, we will show that the aESCT in (\ref{aesct}) is compatible with gauge symmetry, \textit{i.e.}, for any gauge transformation $\mcal G$ as defined in (\ref{91}), the aESCT operator satisfies 
\begin{align}
\dt{\delta^\sigma}\, \mcal G A^\mu = \mcal G\, \dt{\delta^\sigma} A^\mu.
\label{true}
\end{align}

To see this, let us consider an arbitrary gauge configuration $\mcal G A^\mu$. The action of the aESCT on $\mcal G A^\mu$ is
\begin{align}
\dt{\delta^\sigma}\, (\mcal G A^\mu) &= \mcal G\, \mcal G_T^{-1} \,(1+\delta^\sigma) \; \mcal G_T\, \mcal G^{-1} ( \mcal G A^\mu ) - \mcal G A^\mu \nonumber\\
&= \mcal G \big[ \mcal G_T^{-1} \,(1+\delta^\sigma) \; \mcal G_T - \mcal G  \big] A^\mu \nonumber\\
&= \mcal G \,\dt{\delta^\sigma} A^\mu.
\end{align}
Notice that in the first line, the gauge transformation $\mcal G_T \,\mcal G^{-1} $ is performed to bring the gauge configuration $\mcal G A^\mu$ to Lorenz gauge before the SCT is applied. 

Since (\ref{true}) is verified,  the aESCT is thus a valid transformation for Maxwell theory.

Although $\dt{\delta^\sigma} A^\mu$ in (\ref{94}) is distinct from $\widetilde {\delta}^{\sigma} A^\mu$ defined in (\ref{51}), interestingly they have very similar properties; for example
\begin{align}
\dt{\delta^\sigma} F^{\mu\nu}=\widetilde {\delta}^{\sigma} F^{\mu\nu},
\end{align}
the actions of the aESCT and ESCT on the field strength $F^{\mu\nu}$ are identical. This can be readily checked by using (\ref{f}).

Furthermore, the action of the aESCT on the physical field $A^\mu_T$ is the same as $\widetilde {\delta}^{\sigma} A^\mu_T$ defined in (\ref{cor}): 
\begin{align}
\dt{\delta^\sigma} A^\mu_T &= \dt{\delta^\sigma} A^\mu - \frac{1}{\square} \partial^\mu \partial_\alpha [\dt{\delta^\sigma} A^\mu]\nonumber \\
&= K^\sigma A^\mu_T - \frac{1}{\square} \partial^\mu \partial_\alpha [K^\sigma A^\mu_T]\nonumber\\
&= K^\sigma A^\mu_T -d \frac{1}{\square}\partial^\mu A^\sigma_T\nonumber\\
&=\widetilde {\delta}^{\sigma} A^\mu_T,
\label{95}
\end{align}
where we have used (\ref{81}) to get the third equality. 

However, 
\begin{align}
\dt{\delta^\sigma} A^\mu_L &=  \dt{\delta^\sigma} A^\mu - \dt{\delta^\sigma} A^\mu_T\nonumber\\
&= d \frac{1}{\square} \partial^\sigma A^\mu_T \nonumber\\
& \neq \widetilde {\delta}^{\sigma} A^\mu_L, 
\label{96}
\end{align}
where $\widetilde {\delta}^{\sigma} A^\mu_L $ is defined in (\ref{md}).
Nevertheless, since both $\dt{\delta^\sigma} A^\mu_L $ and $\dt{\delta^\sigma} A^\mu_L $ are  pure gauge of $\mcal O(\varepsilon^1)$, they can be related by an infinitesimal gauge transformation. Thus, $\dt{\delta^\sigma} A^\mu $ and $\widetilde {\delta}^{\sigma} A^\mu$ are equivalent up to a gauge transformation.

Therefore, the aESCT defined in (\ref{94}) is equally as good as the ESCT defined in (\ref{full}) for deciding whether a gauge invariant Abelian gauge theory has  ESCT invariance. This is because the difference between the actions of aESCT and ESCT on $A^\mu$ is an infinitesimal gauge transformation, and thus their actions are the same on physical observables.

\end{appendices}

\bibliographystyle{utphys}
\bibliography{conformal}

\end{document}